# The Case for Zero Trust Digital Forensics

Authors:

Christopher Neale[A]

Ian Kennedy[A]

Blaine Price[A]

Yijun Yu[A]

Bashar Nuseibeh[A,+]

[A]The Open University, Milton Keynes, MK7 6AA, United Kingdom
Christopher.Neale@open.ac.uk

[+]Lero, University of Limerick, Republic of Ireland

**Corresponding Author:**
Christopher Neale – Christopher.Neale@open.ac.uk

## Abstract

*It is imperative for all stakeholders that digital forensics investigations produce reliable results to ensure the field delivers a positive contribution to the pursuit of justice across the globe. Some aspects of these investigations are inevitably contingent on trust, however this is not always explicitly considered or critically evaluated. Erroneously treating features of the investigation as trusted can be enormously damaging to the overall reliability of an investigation's findings as well as the confidence that external stakeholders can have in it. As an example, digital crime scenes can be manipulated by tampering with the digital artefacts left on devices, yet recent studies have shown that efforts to detect occurrences of this are rare and argue that this leaves digital forensics investigations vulnerable to accusations of inaccuracy. In this paper a new approach to digital forensics is considered based on the concept of Zero Trust, an increasingly popular design in network security. Zero Trust describes the practitioner mindset and principles upon which the reliance on trust in network components is eliminated in favour of dynamic verification of network interactions. An initial definition of Zero Trust Digital Forensics will be proposed and then a specific example considered showing how this strategy can be applied to digital forensic investigations to mitigate against the specific risk of evidence tampering. A definition of Zero Trust Digital Forensics is proposed, specifically that it is 'a strategy adopted by investigators whereby each aspect of an investigation is assumed to be unreliable until verified'. A new principle will be introduced, namely the 'multifaceted verification of digital artefacts' that can be used by practitioners who wish to adopt a Zero Trust Digital Forensics strategy during their investigations. A qualitative review of existing artefact verification techniques is also conducted in order to briefly evaluate the viability of this approach based on current research efforts.*

# 1. Introduction

Digital Forensics is now a common part of many criminal investigations, and also features in other situations such as corporate incident response resulting from attacks on information systems. Regardless of the context for a digital forensic investigation, it is imperative that it is reliable, a necessity acknowledged by many authorities responsible for the development and regulation of digital forensics who have emphasised the need to formally demonstrate reliability of the methods and tools used within the discipline (Hughes and Karabiyik, 2020). In the UK for example, the Forensic Science Regulator urges all forensics disciplines to embody this by being rooted in 'good science'. This emphasises the need to be transparent about the limits and/or methodology used (Tully, 2020, p.2). One of the phenomena which undoubtedly impacts the reliability of any investigation is trust, yet the role of trust in digital forensics has not been given much attention. Marsh (1994) first tried to formalise trust as a computational concept and studied the application of trust within the interactions between artificial systems. In this work, he acknowledged that there is no accepted definition of trust (something which is still true today), arguing that the presence of trust necessarily implies some degree of uncertainty to an outcome. Given this, while taking the presence of uncertainty to be in conflict with the requirement for reliability and 'good science', it is clear that the role of trust is an issue for careful consideration within digital forensics.

It is perhaps surprising therefore that trust does not play a bigger role in established models of the digital forensics process. In the UK, one of the most widely acknowledged methodologies for digital forensics is that put forward by the UK Association of Chief Police Officers (ACPO, 2012) in their 'Good Practice Guide'. This document outlines the generally accepted principles to be followed by the field, as well as the broad phases of an investigation which are: identification of evidence, collection of evidence, analysis of evidence and presentation of evidence. Interestingly the word trust only appears in one appendix of this guide and only in the narrow context of establishing whether particular forensic tools should be treated as 'trusted'. Part of the reason for this may be the recommendation by the UK Law Commission (The Law Commission, 1997) to repeal without replacement section 69 of the Police and Criminal Evidence Act (PACE, 1984) which required that computer evidence was not admissible unless it was shown that there were 'no reasonable grounds for believing the statement to be inaccurate' or 'at all material times the computer was shown to be operating correctly'. The reason was the significant burden placed on those presenting computer evidence, however the replacement of this position by a presumption that computer evidence is reliable unless shown otherwise may be a part of the reason for the discipline of digital forensics failing to explicitly consider manifestations of the phenomenon of trust within investigations.

This presumption may not last much longer however, as an invited paper by the Law Commission (Ladkin et al., 2020) proposes that this presumption is replaced by a 'third way' where the courts should attempt to actively determine whether particular evidence has been affected in a material way by computer error. In other words, trusting such digital evidence should no longer be the default position, something which may have significant implications on digital forensics practice. However, trust is not entirely absent from the technical literature on digital forensics, as shown by Rekhis and Boudriga (2010) who focus on a particular aspect of trust, namely the potential for the tampering of digital evidence prior to the instigation of a digital forensic investigation. In this work, the authors propose a version of the digital forensics process which includes phases which aim to conduct 'analysis of anti-investigation attacks'. Nevertheless, explicit consideration of trust within the processes used by digital forensics practitioners is generally lacking from the literature, which is an area of concern.

Digital forensics is not the only discipline which has to contend with the challenges presented by the issue of trust. In network security, an increasingly popular paradigm is Zero Trust, a security model and mindset which according the United States National Security Agency (NSA, 2021) allows for sensitive data, systems and services to be better protected from sophisticated cyber threats. This is achieved through design principles and strategies which aim to minimise the reliance of network defenders on trust through enhanced verification of network interactions, such as authorisation events. The National Institute for Standards in Technology (National Institute of Standards and Technology, 2020) have also recently published a standard on implementing a Zero Trust Architecture within networks. The important of Zero Trust to the future of network security is demonstrated by these recent publications, therefore it is worthwhile to consider whether the principles of Zero Trust can be applied more widely, for example, whether they can be used as a basis to begin to meet the challenges presented by trust within digital forensics, as outlined above.

In this paper a new strategy of 'Zero Trust Digital Forensics' is proposed. In particular, the following contributions will be made:
1. A motivating example will be presented and used to highlight the issue of trust in digital forensic investigations.
2. A definition of 'Zero Trust Digital Forensics' will be proposed along with a broad set of ideas of how it could be implemented in practice to provide enhanced verification of digital artefacts.
3. A qualitative discussion of current digital artefact verification techniques which could be applied to the motivating example is presented which is then used to highlight gaps in current research.
4. Further areas of future research will be established.

The rest of this paper is structured as follows. In section 2, a motivating example featuring a digital forensic investigation of an Unmanned Aerial Vehicle (UAV) is outlined, with specific risks relating to trust discussed. Section 3 provides a brief review of current literature which focus on the broad role of trust in digital forensics, before considering one specific challenge associated with this trust, namely identifying whether digital artefacts have been tampered with prior to an investigation being initiated. Section 4 will first outline Zero Trust as a strategy as applied to network security, then consider how these principles could be applied to digital forensics. A definition of 'Zero Trust Digital Forensics' is proposed and considered specifically in the context of identifying artefact tampering. Section 5 provides a discussion of the consequences of the work presented in section 4, alongside an outline of how Zero Trust Digital Forensics could be used to meet the challenges presented in the motivating example Additionally, a brief qualitative review of available digital artefact verification techniques is then conducted which in turn highlights gaps and future research to be addressed. Finally, section 6 concludes with a summary of the paper's contribution.

## 2. Motivating Example

In this section, a simple motivating example will be described and used as a vehicle for discussion throughout the rest of the paper. The intention is to provide a means of highlighting where trust may occur in an investigation so that this phenomenon can be more easily scrutinised.

During the course of an investigation into a serious crime a particular suspect becomes the primary focus of the investigating officers. During questioning, the suspect claims that at the time in question, they were at home, asserting also that another member of their household was playing with a UAV and using it to take pictures which may potentially have captured them during this period, proving their alibi. A UAV is located at the suspects home and seized for further investigation. Information is successfully retrieved from the device and submitted for further analysis. The device in question has one camera which is able to take pictures but no video capture facility. The digital artefacts that can be analysed include images in the 'Joint Photographic Experts Group' (JPEG) format and several files in a proprietary format (PUB – no confirmed acronym) which on initial inspection appear to be some kind of flight log.

An example artefact can be seen in Figure 1, a JPEG image file retrieved from the UAV and Figure 2, which is the output of a tool that extracts 'Exchangeable Image File Format' (EXIF) information from JPEG images and has been run over the image in Figure 1.

*Figure 1 – the picture retrieved from the UAV appearing to show the suspect sitting on a black sofa.*

*Figure 2 – partial output from a tool used to extract EXIF information from the image in Figure 1*

For the purpose of illustrating the role of trust in this investigation, some relevant assumptions which may be present in the process so far are provided in Table 1.

| Assumption | Entities involved |
|---|---|
| That the digital artefacts have been accurately and completely retrieved from the device | Forensic tool(s) and investigator operating system (OS) <br> Suspect system <br> Investigator |
| That the digital artefacts are correctly interpreted on the forensic system | Forensic tool(s) |
| That the digital artefacts have not been tampered with at any stage (i.e., prior to retrieval, during the recovery process or after the data is in custody) | Suspect and other 3rd parties with access to the device (including the investigator themselves) |
| That the investigators interpretation of the digital artefacts is sound | Investigator |
| That the investigators methods (e.g., choice of tool) are valid | Investigator |

| That the system is behaving 'correctly' such that no significant errors exist | Suspect system |
|---|---|
| That the integrity of the storage medium remains for the duration of the investigation (i.e., it does not degrade such that data is altered) | Storage medium |
| That any assumptions made about related systems / environment hold true | Related systems / environment |

*Table 1 – description of potential assumptions in the motivating example*

Clearly the phenomenon of trust has some relevance to these assumptions and therefore plays some role in the digital forensic investigation process. This concept is not unfamiliar to practitioners and some existing standard practices will seek to mitigate specific aspects. For example, it is common to use multiple tools to do the same analysis (often referred to as 'dual tool verification' as promoted by ACPO, 2012) in order to provide some level of verification of the accuracy of each tools output. Whilst imperfect, this does in some way address the issue of trust in any single tool, although it does not guarantee an improvement in reliability (Horsman, 2019). Difficulties may arise with this specific scenario however if common forensic tools don't support the model of UAV and if there is little or no published forensic research into it, both of which happen to be true. Similarly, digital investigators can undertake certification, such as the Certified Digital Forensics Examiner (CDFE) offered by National Initiative for Cybersecurity Careers and Studies (NICCS, 2021), in order to demonstrate their ability to correctly analyse digital artefacts. However, in this scenario it is quite possible that the investigator (and others within their local community on whom they may rely for ad-hoc support) has no previous experience with this device.

There are also additional shortcomings which are present in this case. The amount of trust that it is appropriate to place in any given suspect for example is less straightforward to assess than the competence of practitioners. This would suggest that some means to verify the integrity of the digital artefacts is required. Additionally, in this example, as the suspect system is not a widely used device there is a danger that assumptions made about the related systems and/or environment could be false. By way of example, this includes any assumptions made about the capabilities of the UAV and any potential hardware or software modifications that could have been made to it.

In summary, there are two key issues which require further attention. First, given that the phenomenon of trust is clearly an important factor which influences digital forensic investigations, it is important to identify when this occurs. The assumptions listed in Table 1 are the result of an informal qualitative evaluation of the presented hypothetical scenario and is therefore unlikely to represent a comprehensive assessment of where trust will occur in more complex situations. Secondly, in aspects where it has been established that trust is a factor, it is also imperative to establish the appropriateness of such trust. The danger of not doing this is obvious and has the potential to result in inaccurate, unreliable and risky digital forensic investigations.

# 3. Literature Review

In this section, a brief review of related literature will be conducted. First, the role of trust in digital forensics will be examined which will show that the phenomenon of trust is fundamentally linked to the processes involved in producing reliable digital forensics, despite the lack of attention in current research. Then, some common challenges experienced by practitioners trying to conduct reliable digital forensics will be examined and the specific aspect of potential tampering of digital artefacts proposed as a useful example of an aspect of digital forensic investigations that is contingent on trust. It will also be shown that identifying such activity is an open problem and therefore is worthy of further research.

## 3.1 The role of trust in digital forensics

Trust is a phenomenon experienced by all and is a fundamental part of the human experience, providing a means for understanding and adapting to the complexity of our environment in the face of uncertainty (Marsh, 1994). The potential dangers of trust in the digital domain were identified by Thompson (1984) who provides an example of a malicious C compiler which contains a trojan horse, and who concludes that the moral of such an example is obvious in that only code that has totally been created yourself can be considered trustworthy. In a reflection of this work in Spinellis (2003), the author also concludes that given the unbroken track records of failed security technologies, any claim of a systems trustworthiness should be viewed with scepticism. Marsh (1994) sought to formalise the concept using mathematical notation so that it could be better understood and studied, particularly in the context of artificial agents, such as Distributed Artificial Intelligence. Furthermore, Marsh argues for the utility of assigning numerical values to trust as a way of better assessing and comparing trust in different situations. Several different instantiations of trust are described, namely basic trust (derived from past experience in all situations which can be likened to a general disposition to trust), general trust (trust in an individual entity) and situational trust (trust in an individual entity in a specific situation). The context provided by digital forensic investigations is specifically concerned with this third variation, as this provides the most precise and detailed consideration of trust which can be applied to the setting of digital forensics. Marsh (1994) provides a formula for calculating suitable values for situational trust, which is the product of three combining factors: the utility of the situation (how much knowledge can be gained), the importance of the situation (a subjective judgement of relevant to the originating agent) and general trust, which is modelled as a constant value. How this constant value can be quantified is the subject of discussion within the work, with several suggestions for appropriate strategies provided. Outlining each of these it outside the scope of this paper, however in most cases, one of the contributing factors is based on the history of previous scenarios between the trusting entity and the trusted entity, for example, where previous similar situations have provided positive outcomes for the trusting entity, the value of 'general trust' is likely to be higher than if those previous situations had resulted in negative outcomes.

It is also important to note that situations involving trust are a subclass of situations involving risk (Marsh, 1994) and so it follows that whenever a discussion is had on the role of trust in a given situation (i.e. from one entity to another), this is necessarily related to the risk being taken by the trusting entity. Furthermore, it can be argued that by quantifying this situational trust as encouraged by Marsh (1994), the accuracy of such quantification also has an influence on the amount of risk being undertaken by the trusting entity. By way of example

Marsh (1994) discusses the situational trust he places in his brother, whereby he assigns a high value of trust in his brother to drive them both to the airport, but much less trust in his brother to fly the plane. Clearly a miscalculation of these values could lead to the trusting entity placing themselves in situations of very high risk indeed! For the discipline of digital forensics, questions surrounding risks taken also impact heavily on questions of reliability, which as highlighted earlier, is considered to be a fundamental requirement of the scientific process. Considering the motivating example described in section 2, the manifestations of trust described in Table 1 could also therefore also be thought of as the start of a risk assessment which could be applied to the investigation, where the analysis of each risk equates to determining an appropriate calculation of trust the investigation should have in the particular aspect being considered.

A real-life example of misplaced trust in digital artefacts can be found in the case of Bates & Ors v the Post Office Ltd (No 6: Horizon Issues) (2019). This case was concerned with one of the largest miscarriages of justice ever experienced in the UK, where many sub-postmasters were prosecuted for theft and false accounting (among other charges) based on data collected from the 'Horizon' IT system. The Bates v Post Office case made public information relating to this system which illustrated how the data produced by Horizon could be corrupted and inaccurate, for example due to the fact that it could be accessed remotely which had previously been denied. This complex case demonstrates the danger of misplaced trust in digital artefacts which until this case, had sadly resulted in many lives of innocent people being ruined due to prosecutions of crimes that they had not in fact committed.

Recently Casey (2019) studied the integration of digital forensics into forensic science more generally, stating that 'some practitioners still view the discipline as factual and not requiring scientific treatment.', adding that the uncertainty in digital traces may not be obvious, requiring careful study and experimentation to observe and explain. While not using the term trust specifically, it is clear that Casey (2019) recognises the important role that trust plays, expressing concern about how practitioners may deal with the inevitable uncertainty that arises naturally, however he does not go as far as describing trust explicitly, not mention any attempts to quantify it in any way. Another study by Reedy (2020) focussed on reviewing digital evidence presented over the course of a four-year period, concluding that the impact of human errors and human fallibility were found to be widespread within the discipline. While the author states that digital forensics is becoming increasingly sound with regards to the application of the scientific method, the importance of the human role in digital forensic processes inevitably leads to such circumstances. In a similar way to Casey (2019), while trust was not the focus of the study, it is clear that trust, particularly when incorrectly calculated or assigned, is a contributing factor to such errors. In the case study from section 2 for instance, erroneous trust placed in the accuracy of digital forensic tool used for extracting the evidence from the UAV would be an example of the sorts of errors that both Casey (2019) and Reedy (2020) discuss. These errors are potentially difficult to observe and explain, adding to the overall uncertainty of the investigation and undermining the soundness of the scientific method.

It is also worth expanding on this point a little further, namely the relationship between how trust may be applied to different aspects of a digital forensic investigation, and trust in digital forensics by external parties, such as jurors, witnesses and victims. The latter here is concerned with the perceived reliability of the investigation by external agents, whereas the former is a factor in determining this reliability. According to Arshad et al. (2018), a lack of trust in the digital forensic process gives a smooth and accessible path for defence attorneys to challenge digital evidence which in turn emphasises the need to prove the domain as a rigorous, pragmatic and reproducible science. There is therefore a danger that these issues can impact not only the situational trust in digital forensics (i.e., trust in a specific investigation), but also general trust (in a specific practitioner or laboratory) and in the worst case, basic trust in digital forensics. An interesting yet significant problem for the discipline of digital forensics is understanding how an investigation such as the motivating example from section 2 can be conducted in a way that is demonstrably reliable alongside the inevitable uncertainty that will be experienced through factors such as unproven forensic tools and the potential for artefact tampering to have occurred.

The literature therefore seems to be hinting at an inverse relationship between the trust by digital forensic investigations and the trust in digital forensics by external parties. This paper suggests from this that the more an investigation misjudges appropriate values for situational trust through over-estimation, the more risk it takes. The more risk it takes, the less reliable it becomes, reducing trust placed in it. It makes sense therefore that if increasing trust in digital forensics is a desired outcome, then an obvious strategy is to decrease the trust used by digital forensic investigations, by means of appropriate quantification and additional verification.

### 3.2 Common challenges in conducting reliable digital forensics

There have been several studies which have focussed the variety of challenges in conducting reliable digital forensics. One such example in Lyle (2010) attempted to establish the reasons behind why many digital forensic tools don't have established error rates, unlike many of the tools used in other physical forensic disciplines, when this is considered one of the fundamental requirements of so-called 'good science'. Garfinkel (2010) also included error rates when looking at the main digital forensics challenges at that time, whilst also pointing out that forensic tools should improve their ability to detect and present outliers to human investigators or other artefacts which are seemingly out of place. Furthermore, he states that the inability to extract information from devices in a repeatable manner means that those devices are unable to be analysed for malware. While some of these challenges have been addressed in the past 10 years, for instance the ability of forensic tools to extract data from mobile devices; more recent studies have shown that there are still some current problems relating to reliability. This is not only a human issue and can also be the consequence of technical limitation, such as the implementation of cryptographic protections on systems by manufacturers. Arshad et al. (2018) suggest that more work is needed by new digital forensic techniques in order to verify their accuracy using systematically tested methodologies, while Casey (2019) laments the lack of integration between digital forensics and physical forensics, citing this as a reason for the perceived unreliability of the discipline. All of these challenges are present in the motivating example described in section 2. For example, it can be reasonably assumed that the device is not supported by common tools. The UAV in question is produced by a well-known manufacturer, but as the model is one of several cheaper 'toy' models, has not received any attention from commercial tool suppliers or academic research. Tools exist to extract data from other models of UAV made by the same company, and when tested these may even appear to function and produce potentially plausible results. However, it is unclear how a practitioner can be sure that this process is not producing the sorts of systematic errors that Lyle (2010) refers to. Additionally, determining whether malware is present, as considered by Garfinkel (2010), is non-trivial with no standardised method or process that can be undertaken. Considering the issue argued by Casey (2019), it is unlikely that this type of situation would be accepted in physical forensic disciplines where a comparison could be made to the use of untested and uncalibrated tools. These challenges are potentially uncomfortable to investigators and could potentially lead defence attorneys to the 'smooth and accessible path' to challenge any findings as feared by Arshad et al. (2018).

One particular challenge is the fact that the digital artefacts recovered and analysed during a digital forensic investigation can be tampered with before an investigation starts, meaning that any findings made as a result may be compromised or inaccurate (Arshad et al., 2018; Casey, 2019). This is often referred to as 'anti-forensics' in the literature and continues to be a major challenge to the field. Harris (2006) first tried to establish a common consensus on what this term meant, defining it as 'any attempts to compromise the availability of usefulness of evidence to the forensics process'. Clearly artefact tampering is fundamentally linked to the issue of trust within digital forensics as assessing appropriate values of trust for such artefacts would necessarily require an understanding of whether they have been subjected to such activity. In the example presented in this paper, there is a clear incentive for the suspect to conduct artefact tampering for the purposes of providing them with a plausible alibi, yet it is non-trivial for any practitioner in this situation to decide on an appropriate course of action to verify this in any way. Conlan et al. (2016) emphasise the potential dangers, stating that detection of anti-forensic activity was worthy of further research and initiatives. Their paper offered a new taxonomy on the subject, as well as a practical resource to help digital forensic practitioners identify when it might have taken place. However, this was rather limited in nature, consisting simply of a hash database of known anti-forensic tools which could only be used to detect whether such tools were currently present on a particular type of system. Additionally, the studies conducted in both Freiling and Hösch (2018) and Schneider et al. (2020) demonstrate that it is certainly possible to tamper with digital artefacts without this being detected by forensic tools or experience practitioners, although some uncertainty remains about how difficult this is in practice for different types of devices and artefacts. Given the results of these experiments, along with the incentive for the suspect in the motivating example to tamper with artefacts and the challenges of the particular model of UAV described earlier, understanding how likely this or, how difficult it would be to conduct and verifying whether it has indeed happened, are all relevant challenges the investigator is required to solve.

If artefact tampering activity is not identified, it is inevitable that this will result in any investigation producing incorrect outcomes and this can seriously undermine its reliability. If digital evidence cannot be proven to be authentic and reliable, then it is meaningless to present it in a court of law (Yusoff et al., 2010), while Arshad et al. (2018) state that any inability to identify the evasive behaviours present in artefact tampering directly affect the reliability of digital evidence, potentially creating serious doubt in a court. There have been very few studies which have investigated the extent to which artefact tampering successfully defeats attempts at forensic investigation, however one study by de Beer et al. (2015) concluded that in the context of South African investigations, the experience of practitioners in dealing with artefact tampering was very limited and that despite the fact that practitioners rated the value of finding such activity as being very important, they did not routinely make an effort to identify it in practice. In the motivating example, it is certainly plausible to imagine a scenario where an investigating party does not have the time, resources or ability to fully explore the possibility of artefact tampering.

Clearly then, the potential existence of artefact tampering activity is one element of digital forensic investigations from several that is connected with the phenomenon of trust. Given the conclusions of de Beer et al. (2015), it could well be that it is also a common source of misplaced trust and is therefore the reason this paper will study it in more detail in order to consider the wider role of trust in digital forensics.

## 3.3 Shortcomings in identifying artefact tampering

Current literature also demonstrates that the issue identified in section 3.2, specifically the identification of evidence tampering activity, is not a solved problem and therefore it requires further research. For example, Garfinkel (2007) provided details of some common 'anti-forensic' techniques, stating that a major factor behind their supposed success was the limited resources devoted to finding them by law enforcement agencies. Conlan et al. (2016) noted that a significant limitation of their work to provide an extended taxonomy on artefact tampering techniques was the sheer number of tools available to conduct them, calling for further research into anti-forensics to produce and accessible body of knowledge which they feel is likely to be useful to practitioners in aiding them in their attempts at detection. Furthermore Casey (2018) proposes using what he terms 'Digital Stratigraphy' techniques to improve the contextual analysis of digital artefacts during investigations to root out artefact tampering, however he points out the need for tools to provide the relevant information in order for this to be successful. He gives the example of activity such as mass file deletion and file tunnelling as potential examples of the sort of behaviours that common forensic tools do not identify. Bhat et al. (2020) conducted an experiment to find out whether forensic tools can extract complete and credible evidence from digital crime scenes that had been tampered with by file-system attacks. They concluded that the implication of their work was that investigators cannot absolutely rely on such tools due to examples of tools which had failed during their experiments, calling for more research into the common pitfalls of these types of software.

In the absence of adequate tooling being available to digital forensic investigators to identify artefact tampering, some attempts have been made to produce generalised techniques which aim to meet this need. Shanmugam et al. (2011) proposed a technique to formally validate digital evidence in order to detect anti-forensic attacks that may have been conducted by a suspect. The technique uses decision tree analysis to match analyse digital events against known bad attack patterns. However, the technique is limited by the fact that is relies on knowing all possible artefact tampering methods in advance in order to incorporate them into the model. Furthermore, the model isn't proven against complex attacks which make minor modifications to these known patterns. Rekhis and Boudriga (2012) proposed an inference-based rules system where systems were formally modelled using the Temporal Logic of Actions notation. Rules were then specified which could be used to identify attacks on the forensic process due to artefact tampering. However, like Shanmugam et al. (2011) this relied on knowing what all these attacks would look like prior to the start of an investigation (referred to in the latter as a 'library of attacks'). Additionally, the Rekhis and Boudriga (2012) method relies on the presence of sources of digital evidence which are provably secure, alongside a known good initial system state. Both of these conditions are impractical for many real-life investigational scenarios, while both techniques are also only ever tested on a limited type of device and so their wider applicability is unproven. Horsman and Errickson (2019) proposed using signatures of known anti-forensic tools in order to identify their use on suspect systems. These 'Digital Tool Marks', which they refer to as DTMs, allow the practitioner to identify the existence of inconsistencies on a system due to the use of these tools, however the method relies on having an up-to-date database of DTM signatures, something that does not exist at this time. They also do not propose any way of dealing with false positives generated by such signatures or any way of identifying manual tampering that may take place through tools that are typically already present on common systems such as a hex editor on Linux. A further technique proposed by Mothi et al. (2020) uses a mathematical principle for validating phases of a digital forensic investigation through the use of counter tensor products. However, their technique again relies on a knowledge of the different tampering attacks that can take place as well as how they can be countered for every phase of the investigation. Additionally, the technique only informs the investigator when tampering may have occurred and does not provide a means for positive identification of this activity. Given the significant limitations on these techniques, alongside the fact that none of them are in widespread use by the practitioner community, it is argued here that the noble aspiration of a generalised technique to identify artefact tampering is still elusive to the community.

Additional techniques have been proposed which are less generalised in nature, but instead focus on specific types of artefact tampering. Arasteh et al. (2007) for instance proposed a formal analysis method using a custom model checking approach for log files typically generated on a variety of systems in order to detect whether they had been tampered with or generated naturally as part of normal system operation. Unfortunately, no algorithm was publicly provided which could be used to apply this technique to additional scenarios than the case study used in the paper. Rowe and Garfinkel (2012) produced a proof-of-concept custom 'Dirim' tool to analyse file metadata and identify anomalous and suspicious files based on inconsistencies found at the hardware level which could be used to find clues relating to potential metadata tampering of such files. Other approaches could be considered more theoretical than practical, for example Shanableh (2013) outlined a machine learning approach which could be applied to video files in order to detect instances of frame deletion which could indicate temporal manipulation of such files. Like Arasteh et al. (2007), while the mathematical constructs used in the approach were detailed, no publicly available tool was produced as a result, limiting the applicability of this approach to a wider community of practitioners. Pieterse et al. (2018) studied the Android mobile platform and converged on several 'theories of normality' which they proposed could be used to test digital artefacts against in order to detect any inconsistencies which may infer tampering via formal modelling of those systems. There are many more examples of such approaches which exist in the literature on digital forensics, where each has a limited scope of applicability to potentially identify specific types of artefact tampering. However, with the broad range of devices that practitioners are faced with, it remains to be seen whether existing approaches such as these examples would provide adequate coverage for investigators to be confident that artefact tampering could be identified in the majority of cases. This will be considered further in section 4.3. In the motivating example from section 2, there is no academic literature which researches methods for identifying digital artefact tampering on the model of UAV in question.

## 3.4 Summary

It is clear that trust is an important component of digital forensic investigations, however it is rarely, if at all, explicitly considered. It is intimately linked to the overall reliability of such investigations and literature on the subject hints at an inverse relationship: the greater the amount of trust placed in aspects of the investigation (such as the provenance of the digital artefacts), the lesser the amount of trust there can be that the investigation is reliable. Therefore, in order to increase the reliability of digital forensic investigations, it would seem that an obvious approach would be to decrease the amount of trust placed in these features.

This paper therefore proposes a new model for digital forensics which is outlined in the next section. Furthermore, this model will be considered in the context of identifying potential artefact tampering in order to understand how it could potentially be used in practice, as well as provide a qualitative evaluation of the extent to which it would be practical to apply such a model based on currently available tools and techniques.

## 4. A Zero Trust digital forensics model

In the previous section, current literature relating to digital forensics was considered which showed the importance of reliability to the discipline. Trust was shown to be an important component of reliable digital forensic investigations. This section will consider the Zero Trust security strategy which is gaining popularity in network security. To date there has been very little work looking at the application of Zero Trust to the discipline of digital forensics. The only explicit link the current literature as of the time of writing is Mary et al. (2021) who believe that employing a Zero Trust approach to digital forensic investigations could have some benefits. These include reduced time delays in investigations, a reduced risk of ignoring malware and reducing issues connected with reaching incorrect verdicts. However, the authors of this work do not provide any detail as to how Zero Trust could be applied in order to realise these benefits in practice and there is no discussion regarding how manifestations of trust can be identified within investigations. Furthermore, no definition is provided by these authors for Zero Trust Digital Forensics and no examples given to illustrate the utility of the approach. Therefore, this section will examine the main principles behind Zero Trust in more detail before exploring how this could be applied to digital forensics. A definition of 'Zero Trust Digital Forensics' will be given, and a proposal made for how Zero Trust principles could be applied to the specific risk of digital artefact tampering.

### 4.1 Zero Trust as a security strategy

The discipline of network security has traditionally had a strong focus on trust due to the need for decisions to be made by systems for granting access to resources to different network entities. However, in recent times, traditional means for achieving this aim, such as user authentication via passwords have come to be seen as inadequate, and this has led to a move towards a Zero Trust Architecture model to be promoted by several organisations. The parallel with digital forensics is clear. In a similar way to network security, the previous sections have illuminated the need for digital forensics to ascertain which entities, or aspects of investigations, should be considered as trusted and so the discipline therefore can potentially learn from how network security has approached this issue.

In 2020, the United States National Institute for Standards in Technology (NIST) produced a special publication titled 'Zero Trust Architecture' (National Institute of Standards and Technology, 2020). In this publication, the authors assert that organisations should adopt the posture that an attacker is present in the environment and that there should be no distinction in the perceived trustworthiness of assets inside or outside the enterprise network, all should be regarded as potentially hostile. NIST (2020) stipulates that Zero Trust is not in fact a single architecture, but instead a set of guiding principles upon which workflows, systems and operational processes can be designed. Furthermore, the dynamic nature of trust is addressed, with the document specifying that all trust must be explicitly granted and constantly evaluated. Seven tenets of Zero Trust are specified, which further describe how these principles can be practically applied to network architecture, for example 'All resource authentication and authorization are dynamic and strictly enforced before access is allowed'. NIST (2020) also describe the use of an abstract 'trust algorithm', which they define as 'the process used by the policy engine to ultimately grant or deny access to a resource.' In other words, in order to implement Zero Trust, some algorithmic means must be established, the output of which forms the basis upon which decisions can be made about what aspects should be trusted and to what extent. Relating this back to the work of Marsh (1994), the output of such a trust algorithm would be used as the value of 'general trust', one of the factors, along with 'importance' and 'utility' which is required to calculate 'situational trust'.

In the same year, the United Kingdom's National Cyber Security Agency (NCSC, 2020) produced a beta release of their guidance titled 'Zero trust principles'. This includes what they term as the 'un-trusting 8' – overarching principles for the application of Zero Trust. These are broadly in agreement with NIST (2020) and include the principles of 'Authenticate everywhere' and 'Don't trust any network including your own'. As of the time of writing, this publication is only released as beta, and therefore subject to further changes, demonstrating that an understanding of what should be included when framing a discussion on Zero Trust, along with how it should be applied in practice, is still in its infancy.

Additionally, the United States National Security Agency (NSA, 2021) also published a report titled 'Embracing a Zero Trust Security Model' in which they also promote a Zero Trust approach to network architecture. They describe Zero Trust as a set of system design principles which acknowledges a wide range of threats. The main goal according to this document is to eliminate implicit trust in network assets in order to secure systems and to ensure that anomalous and malicious activity is identified at the earliest opportunity. Zero Trust is described as a mindset, with three guiding principles: 'Never Trust, always verify', 'Assume breach' and 'Verify explicitly'.

The fact that these three organisations, all of which have established reputations within the field of network security and none of which are motivated by profit, have recently published specific guidelines on implementing Zero Trust suggests that it is a concept which is increasingly being taken seriously as a way to genuinely enhance network security. However, the fact that none of the documents agree entirely on what these principles are, nor describe the so-called Zero Trust Mindset in a consistent manner, suggests that the understanding and implementation of the concept of Zero Trust is still immature. For example, NSA (2021) does not distinguish between the similar principles, 'Never trust, always verify' and 'Verify explicitly', nor explain any perceived differences. Nevertheless, some consistent themes are present in all three documents, and these will form the basis of the application made in this paper of Zero Trust to digital forensics. These consistent themes being that trust is not assumed from the outset, some verification method is required before trust can be assigned and that trust is a dynamic component which can be changed over time depending on context.

As an aside, it is worth mentioning that the first reference to the concept of Zero Trust has been linked to Marsh (1994). However, in this work, Marsh (1994) treats the value zero as neutral, whereas in the above documents, Zero Trust (i.e., assigning the numerical value of 0 as a measure of trust) is considered as the lowest possible such value. Marsh (1994) instead assigns values of trust within the range -1 to 1 inclusive, however it is trivial to produce a linear mapping from the range used by Marsh to any others, such as [0,1] or the use of percentages for example. For the sake of clarity, this paper refers to Zero Trust in a way consistent with more recent publications, where 0 is considered to be the lowest possible value of trust that can be assigned, which manifests as active distrust, or as an extreme pessimist (as Marsh would describe it).

## 4.2 Zero Trust within the digital forensic process

It has been suggested in previous sections that, like in network security, the phenomenon of trust is deeply integrated with the discipline of digital forensics. The Zero Trust approach which is still somewhat immature within network security, also has been seen to have some potential applications to digital forensics and so this section offers a definition for this concept, derived from the literature review in section 3.

*Definition: Zero Trust Digital Forensics is a strategy adopted by investigators whereby each aspect of an investigation is assumed to be unreliable until verified.*

Three additional observations are offered. First, while this paper proposes this new definition for the term Zero Trust Digital Forensics, the actions and activities that would be used to implement this strategy are not all fundamentally new. Examples include dual tool verification, which has been common practice for some time, tool testing and validation and practitioner certification. Other aspects of digital forensic investigations however have had less attention, including the verification of the integrity of digital artefacts to tampering activity as discussed earlier. By defining the Zero Trust Digital Forensics strategy, it is hoped that the totality of these activities, including areas of both strength and weakness can further the ultimate goal, namely increasingly the reliability of the output of digital forensic investigations, by minimising the underlying trust that these investigations themselves rely on.

Secondly, it is observed that there is much research in the literature that is already aiming to improve the application of Zero Trust Digital Forensics as defined here. Examples include Casey (2018) which aims to improve the contextual analysis of digital artefacts, the work of Bhat et al. (2020) on tool reliability and Yusoff et al. (2010) on output validation. However, it is hoped that by defining the concept of Zero Trust Digital Forensics, some harmonisation can be brought to these types of research output.

The third observation relates to the achievability of the definition. The logical consequence of the definition is that all aspects of a digital forensic investigation need to be verifiable. However, for many aspects, verification techniques may not even exist, may not be precise and where they are, their application necessarily results in some cost, be that financial, time, effort or combination of these. If uncertainty remains after attempts at verification have been made, then this must be accounted for by practitioners. One potential approach for modelling uncertainty in such complex scenarios, like the one presented in the motivating example, is through the use of probabilistic frameworks. This technique has been applied to various problems in digital forensics in recent years, for example Overill and Silomon (2010) who propose their use in quantifying the extent to which digital artefacts support a given hypothesis. However it is worth noting that there are difficulties with applying this approach soundly, as argued by Nagy et al. (2015) who cite the difficulties in assigning accurate and meaningful probability values from previous investigations to given forensic scenarios.

These issues could lead some readers to conclude that Zero Trust Digital Forensics is simply impractical for application to practice, however this paper seeks to challenge this notion. Certainly, challenges exist in understanding how the theoretical definition can be applied, such as whether the strategy is too financially and computationally expensive for typical investigations, whether adequate verification techniques exist and can be effectively designed, and whether any remaining uncertainty can be properly understood and explained correctly to a wide range of stakeholders. This practical application of the theoretical definition of Zero Trust Digital Forensics necessarily requires further research, but such work is out of scope for this paper.

As an aside Zero Trust Digital Forensics is here described as a strategy rather than a model to be followed by practitioners. The concepts behind it which were discussed in section 4.1 are instead intended to be incorporated into every part of an investigation, regardless of the specific process or methodology being followed. As a result, Zero Trust Digital Forensics does not conflict with existing ways of working, but borrowing a concept from software engineering, can be thought of as a 'wrapper' for existing processes.

At this stage, the motivating example from section 2 is again considered. Each manifestation of trust that was identified in Table 1 has been again recorded in Table 2, however an additional column has been added to provide a suggested method for verification of the aspect of the investigation. Note that this is considered here to be equivalent to the 'trust algorithm' (National Institute of Standards and Technology, 2020) however the means of verification has been expanded to include methods that are not strictly algorithmic in nature.

| Assumption | Entities involved | Possible method(s) of verification / Trust Algorithm |
|---|---|---|
| That the digital artefacts have been accurately and completely retrieved from the device | Forensic tool(s) Suspect system Investigator | Tool testing research, dual tool verification, certification and competency testing |
| That the digital artefacts are correctly interpreted on the forensic system | Forensic tool(s) | Tool testing research, dual tool verification |
| That the digital artefacts have not been tampered with at any stage (i.e., prior to retrieval, during the recovery process or after the data is in custody) | Suspect and other 3rd parties with access to the device | Artefact verification techniques, chain-of-custody procedures |
| That the investigators interpretation of the digital artefacts is sound | Investigator | Certification and competency testing, suitable collaboration, supervision and accountability of investigator |
| That the investigators methods (e.g., choice of tool) are valid | Investigator | Certification and competency testing, suitable collaboration, supervision and accountability of investigator |
| That the system is behaving 'correctly' such that no significant errors exist | Suspect system | System verification techniques |

| That the integrity of the storage medium remains for the duration of the investigation (i.e., it does not degrade such that data is altered) | Storage medium | Storage medium integrity verification techniques |
|---|---|---|
| That any assumptions made about related systems / environment hold true | Related systems / environment | Suitable collaboration, accountability and accountability of investigator |

*Table 2 – description of the role of trust in the motivating example with additionally details potential trust algorithms*

However, these verification methods are still only described at a high level and on their own provide limited progress towards achieving Zero Trust Digital Forensics. Nevertheless, it is suggested that explicitly recording these is a first step in increasing the reliability of an investigation by demonstrating identified areas of trust along with a potential means for verification.

However, in order to provide some further depth to the understanding of Zero Trust Digital Forensics, the next section will look specifically at one area of trust, namely the potential tampering of digital artefacts.

### 4.3 Zero Trust Digital Forensics in relation to identifying artefact tampering

In this section, the potential application of Zero Trust Digital Forensics will be considered in relation to the specific issue of potential artefact tampering. Table 2 identified the manifestation of trust as being 'That the digital artefacts have not been tampered with prior to retrieval', the entity in which trust is being placed as 'the suspect and other 3$^{rd}$ parties with access to the device' and the method of verification as being 'artefact verification techniques'.

One approach to this could be an application of the technique proposed by Rekhis and Boudriga (2012). Figure 3 shows the digital forensic process proposed by these authors in order to counter 'anti-forensic' attacks.

*Figure 3 – Digital Forensic Process proposed by Rekhis and Boudriga (2012)*

Note that there are too many types of so-called anti-forensic attacks to describe in detail here, however the taxonomy in Conlan et al. (2016) presents a recent overview for the interested reader.

An inherent weakness of the inference-based rules approach from the Rekhis and Boudriga (2012) paper is that it requires an understanding of what artefact tampering looks like in order to successfully complete the 'Searching for anti-forensic attacks' and 'identification of all affected evidences' sub-phases. This may work for system types which have been the subject of much research, but for our motivating example this hasn't been the case and so it is difficult to see how such a technique could be applied. Furthermore, similar weaknesses are present in Shanmugam et al. (2011) and Mothi et al. (2020) where in both cases the same requirement of knowing all possible tampering attacks ahead of time is present, while in Horsman and Errickson (2019) where signatures for all potential tools used to conduct tampering need to be created and stored in the database of Digital Tool Marks.

Dismissing existing generalised processes which can be used for verifying the digital artefacts in the motivating example, consideration is instead given to what is needed in order to meet the definition of Zero Trust Digital Forensics for artefact tampering. First, each artefact is assumed to be completely unreliable, at which point it is then subjected to one or more verification techniques. These techniques are likely to be more specific (for example Shanableh, 2013 to analyse video files or the Arasteh et al., 2007 method for checking log files) and so additional thought is needed to understand how and when to apply multiple techniques. For instance, if a proposed technique is able to verify the temporal aspects of an artefact such as a digital video, this does not necessarily mean it is completely tamper-free as it could have been manipulated to appear to have been filmed at a different physical location through tampering of GPS co-ordinates. However, each verification technique does enable practitioners to start to build a 'trust history' as described by Marsh (1994) which can be used to make a more informed choice about suitable levels of trust which can be assigned to the artefact in question.

The above necessitates therefore a discussion about the 'integrity' of digital artefacts where in this case 'integrity' is used to specifically mean an absence of tampering. This can be broken down further as shown in Table 3.

| Integrity sub-type | Questions which enable verification |
|---|---|
| Temporal Integrity | Is the temporal metadata associated with the artefact plausible and consistent with itself and other available artefacts? |
| Syntactic Integrity | Is the structure of the artefact plausible? Can a distinction be made between artefacts created as part of 'normal system behaviour' and those created by an external source such as a human or tampering tool? |
| Semantic Integrity | Is the inferred meaning of the artefact plausible? Is it consistent with other artefacts that are available? Does the 'story' it tells make sense? |

*Table 3 – Identified subtypes of artefact integrity*

Establishing artefact integrity is therefore a multifaceted problem and so in order to follow a Zero Trust Digital Forensics strategy to establish such integrity, there is a need to conduct multifaceted verification of the digital artefacts. The challenge therefore for practitioners is therefore identifying existing techniques which can contribute towards this goal of multifaceted verification, while the challenge for the research community is in designing such techniques for different types of artefacts.

## 5. Discussion

This paper has presented an initial outline of Zero Trust Digital Forensics as a strategy for conducting digital forensics investigations. Common digital forensic methods do not explicitly consider the phenomenon of trust and have an inconsistent approach to dealing with its consequences as shown by the way in which forensic images are subject to chain-of-custody procedures and treated as untrusted until verified, whereas the digital artefacts themselves are often trusted by default. The aim of Zero Trust Digital Forensics is to increase the reliability of such investigations by identifying where trust manifests and then treating all of these features as unreliable until one or more verification techniques has been applied to provide a defined level of assurance. This section explores the consequences of such a definition.

## 5.1 Consequences of Zero Trust Digital Forensics as a general investigative strategy

Zero Trust Digital Forensics requires all parties involved in investigations to identify and evaluate the role of trust in all aspects of their work explicitly. In some cases, this is already a well-understood concept, an example being the ability for digital data to change as a result of forensic actions. Chain-of-custody procedures, such as the hashing of digital images ensure that trust in the integrity of forensic images is well-placed due to this type of verification. Other aspects of investigations typically consider the phenomenon of trust less, one example being the potential for digital artefacts to have been tampered with. For this aspect, verification of these artefacts may only ever take place if practitioners notice particular inconsistencies present within the artefacts. Failure to conduct such verification can have devastating consequences, as demonstrated by Bates v Post Office (2019). Zero Trust Digital Forensics therefore requires the explicit identification of trust in all aspects of investigations, but identification alone is unlikely to substantially increase reliability.

Secondly, the definition, and its implication for the need to verify all aspects of investigations highlights shortcomings in many tools and techniques. As shown in section 4, many artefact verification techniques produced as a result of academic research are not then made available for practitioners in the form of standalone tools or additional functionality to larger tools. Identifying trust and evaluating its role in investigations provides limited benefit to practitioners unless they are then able to conduct remedial action as verification. To become a reality however, practitioners need to be able to easily identify suitable verification techniques, something which is non-trivial given the scale of digital forensics research, and then apply those techniques to real cases.

Thirdly, the practical application of Zero Trust Digital Forensics as a holistic strategy for investigation requires further study. The strategy as presented in this paper is purely theoretical, and understanding its practical application is left for future work but it is argued here that this is a crucial next step. A theoretical strategy in isolation is of minimal utility for the community, but this paper also argues that applying this theoretical strategy based on current research is non-trivial and therefore for it to start to become useful, more research is needed into both the wider application of current verification techniques, as well as into new and more efficient verification techniques. Additionally, following a Zero Trust Digital Forensics strategy necessarily requires an increase in resources from current processes, which can exhibit as time, money, computational and so on. In a world where such resources are finite, an understanding is needed of the scale of these additional means in practice and how they can be minimised, for example through automation. As it stands, due to the limitations of applying the strategy already identified, it would be easy to dismiss Zero Trust Digital Forensics as impractical, but this should instead be a motivation for further work. As a community, the goal that is attained by applying Zero Trust should inspire and motivate research into ways of making such a strategy realistically attainable. Examples of the sort of research proposed, for the specific case of artefact verification, may include a determination of the appropriateness and effectiveness of the principle of multifaceted verification in specific scenarios. Examples include understanding whether a minimal number of artefacts is required in order to provide verification, or whether all artefacts require the same amount of verification in different circumstances. Whether verification can ever be complete is also an important issue for practitioners, alongside establishing suitable error rates for new verification techniques. Applications of existing technologies, such as machine learning and zero-knowledge proofs could potentially also be used in supporting artefact verification. Avenues such as these highlight the richness and potential gains that could be achieved with further research into the application of Zero Trust Digital Forensics.

## 5.2 Current viability of Zero Trust Digital Forensics in relation to identifying artefact tampering

In this section, the specific aspect of identifying artefact tampering will again be the focus. The motivating example is considered alongside the principle of multifaceted verification of digital artefacts stated in section 4.3. There are two artefact types in scope for the investigation as described previously, an image file previously shown in Figure 1 (including the EXIF metadata an excerpt of which is shown in Figure 2) and proprietary PUB files which appear to contain flight information data. Consideration is then given to how the principle of multifaceted verification of digital artefacts can be applied.

First, let's consider the image file, which is in the JPEG format. Table 4 presents a qualitative evaluation of potential existing verification techniques as found in the digital forensics literature.

| Reference | Technique description | Facet verified | Limitations |
|---|---|---|---|
| Counter-forensics: Attacking image forensics<br><br>(Böhme and Kirchner, 2013) | Various statistical tests to detect image manipulation/forgery | Syntactic | Requires knowledge about the conditional probability distributions of images which are difficult to determine accurately.<br><br>Difficult to sample both authentic and counterfeit images efficiently which parts of this technique relies on<br><br>No tool provided; mathematical constructs would need to be manually re-created |

| Anti-Forensics: A Practitioner Perspective (de Beer et al., 2015) | Detection of the use of steganography using known steganalysis tools | Syntactic | Will only find known steganographic techniques/malware for which there are signatures |
|---|---|---|---|
| Forensic Similarity for Digital Images (Mayer and Stamm (2020) | Custom technique called 'forensic similarity' to determine whether two image patches contain the same or different forensic traces via deep-learning methods | Syntactic | Requires an original source for comparison of the artefact in question

No tool is provided to apply these techniques, so would need to be manually re-created |
| Can we trust digital image forensics? (Gloe et al., 2007) | Statistical tests to detect image resampling

Statistical analysis of sensor noise to identify digital camera image origin | Syntactic

Semantic | Neither of the presented techniques is effective in all cases as of the time of the paper

No tool is provided to apply these techniques, so would need to be manually re-created |
| Fighting Fake News: Image Splice Detection (Huh et al., 2018) | Unsupervised machine learning technique to determine the consistency of a JPG based on the files metadata | Syntactic

Semantic

Temporal | Not well suited to finding smaller splices in images

Under-exposed and over-exposed regions are sometimes flagged as inconsistent which is a false positive

Some manipulations (e.g. copy-move) can't be detected as the exact same image source is used and therefore no inconsistency is introduced by this manipulation |

| | | | Model takes several weeks to train and no details are provided as to how practitioners could access the model |
|---|---|---|---|

*Table 4 – potential artefact verification techniques which could be applied to the JPEG image in the motivating example*

Whilst Table 4 is only a small sample of the existing techniques for verifying JPEG files, some initial observations can already be made.

First, it would be extremely time-consuming for any practitioner to implement some techniques due to the lack of available tools created. This suggests that more effort should be dedicated by the research community to ensuring that the output of their work is able to be used in real-life scenarios. Complicated techniques based on abstract mathematical constructs, such as in Böhme and Kirchner (2013) or those based on advanced machine learning techniques such as Huh et al. (2018) are extremely limited in how they can be applied unless this is addressed.

Secondly, the fact that most techniques focus on syntactic integrity and fewer on semantic and temporal (for which there is only one) demonstrate that there is a challenge in ensuring that the principle of multifaceted verification is fully adhered to. For instance, the technique proposed by Huh et al. (2018) is concerned with splice detection and does not account for other tampering that could occur, such as manually adjusting values within the EXIF metadata. It is also possible that some techniques may 'overlap' and understanding where this is the case is essential in ensuring that precious resources are not wasted on re-verification where this has already been sufficiently established.

Thirdly, where there are multiple options for techniques to apply, there are no easy means for comparison. For example, when deciding between Böhme and Kirchner (2013) and (Mayer and Stamm (2020) in order to verify the syntactic integrity of a JPEG image, a complete understanding of each technique is required, alongside a suitable critique of its potential limitations before deciding between them. For practitioners, this is likely to become impractical when faced with the sheer volume of different artefact types encountered. Evaluation of the techniques against standardised datasets would begin to resolve this issue, however the lack of such datasets is already a significant challenge to the discipline, as observed by Arshad et al. (2018).

Finally, the act of simply identifying these potential techniques is time consuming and requires a good working knowledge of current research literature in digital forensics, which itself is vast. As it stands, it is unreasonable to expect individual practitioners to be able to identify relevant research which addresses issues of artefact verification in every case without some means of assistance, for example through an accessible database which captures this information as research is published and disseminated within the community.

Regarding the PUB file, as far as has been established by the research conducted in this paper, no techniques have been published which can be used for verification of this artefact. This highlights a further issue, namely the coverage that existing verification techniques provide for the different types of artefacts that may be encountered by practitioners.

In order to follow a Zero Trust Digital Forensics strategy, an investigator would need a verification method for both the PUB file and the JPEG image in question before either would be considered reliable. If a generalised technique existed, that was accessible for practitioners to find and apply to their cases, which provided this verification, or at a minimum, partial verification alongside enough information to determine which features of the artefacts could not be verified, then this would open the possibility for a Zero Trust Digital Forensics strategy to be realised. However, as seen in section 3, existing generalised techniques contain too many drawbacks for them to be used in such scenarios (such as having a requirement to know about all possible attacks before an incident occurs) and therefore further work is required into developing new techniques. If enough specific verifications techniques existed for both types of files, then these could be applied as a way of following the strategy. However, for the JPEG file, it is currently very difficult to identify, assess and apply existing techniques, which may cover different aspects of artefact integrity or overlap in terms of the assurance they provide. This is due to the fact that they are published in disparate locations with no standardisation of how they are presented to the community and are often lacking in methods of application such as tools plugins or proof-of-concept standalone tools. For the PUB file, verification techniques would need to be designed and evaluated from scratch as none would appear to currently exist. Therefore, it would appear prudent to provide some means for standardising and publishing the output of research into verification techniques in a way that practitioners could easily access these techniques and compare between them. A further improvement would be to ensure that such research, wherever possible, results in publicly available tools for use in practice, whether this be in standalone format or as plugins for more common tools.

## 6. Conclusion

Many areas of technology have had to come to terms with understanding the role of trust and digital forensics is no exception. Failure to properly appreciate the role of trust can have devastating consequences, potentially undermining the reliability of investigations and ultimately the trust that external parties such as the courts can have in their findings. One approach that has been adopted by network security is that of Zero Trust, but the principles contain within this approach can be extended to other disciplines. This paper presents a proposed definition of Zero Trust Digital Forensics, demonstrating its utility through exploring its potential application to a motivating example and outlining some shortcomings that need to be addressed before the strategy can be fully realised.

As a result of this work, it is hoped that the research agenda can be influenced to take more consideration of how Zero Trust Digital Forensics can be applied in practice. In the same way that the term 'forensic readiness' is used by the community to describe research which aims to help organisations prepare for an event requiring a digital forensic investigation, so the provided definition of Zero Trust forensics can help describe research which aims at providing verification of aspects of investigations such that the reliance of trust in these aspects is eliminated. It is expected that new and creative means for providing this verification, whilst working within common constraints such as limited time, computing resource and money will result in greater adherence to Zero Trust Digital Forensics.

Specifically, future research should be concerned with examining the practical application of this strategy to realistic investigative scenarios. This will provide a broader understanding of whether digital forensic practice can be adapted to provide an improved benefit to society through the application of a Zero Trust Digital Forensics strategy. It is anticipated that this could take many forms, and so it is not considered beneficial to provide an exhaustive list of future research here, however examples would be expected to include issues of practicality (such as whether a proposed verification algorithm can perform quickly enough over large sets of digital artefacts), issues of accuracy (such as whether a proposed verification algorithm correctly identified tampered artefacts) and issues of applicability (such as whether enough verification methodologies are available to practitioners).

## Funding

This work was partially supported under SFI grant number 13/RC/2094_P2 and UKRI/EPSRC grant number EP/R013144/1.